# Developing an Open-access Telegram Bot for Automated, estimation and IA-based interpreted of OpenQuake Ground-Motion intensity measures

Boumédiène DERRAS[1,2]

[1]Civil engineering & hydraulic department, university of Dr Moulay Tahar, Saida, Algeria. boumediene.derras@univ-saida.dz. orcid.org/0000-0002-3979-7083

[2]Risk Management and assessment (RISAM) laboratory, University of Abou Bekr Belkaid, Tlemcen, Algeria, boumediene.derras@univ-tlemcen.dz

## 1 ABSTRACT

The prediction of Intensity Measures (IMs) using Ground-Motion Models (GMMs) is a fundamental component of seismic hazard assessment. However, the best estimation of IMs traditionally requires specialised searchers, programming expertise, and the manual sourcing of regression coefficients. This complexity creates a significant barrier to rapid, scenario-based analysis for engineers, researchers, and students. To address this challenge, we have developed the OpenGSIM bot, an automated and conversational workflow that provides on-demand access to the entire OpenQuake-engine GMM library via the Telegram messaging application. The system leverages the n8n workflow automation platform to orchestrate a user interaction, guiding the selection from over 21 IMs, along with their associated total standard deviation (aleatory variability), and more than 500 distinct GMMs classes. The bot interactively elicits the necessary physical parameters, accounting for source, path, site, geological, tectonic, and regional effects as required by the chosen model. A Python wrapper executes the calculation and generates a comprehensive report. A Large Language Model (Gemini) is used to provide an automated interpretation of different types of response spectrum. By abstracting the complexities of software installation and programming, OpenGSIM democratises access to advanced seismic hazard tools, enabling robust IM estimation on any device and without the installation of any software or code package.

**Keywords :** Ground-Motion Model, n8n Workflow Automation, Telegram Bot, OpenQuake, Gemini Large Language Model

# 1. Introduction

The seismic hazard assessment stands as a cornerstone of earthquake engineering (Kramer, 1996). At the heart of this endeavour lies the use of Ground Motion Models (GMMs). Functional forms that predict the ground motion intensity measures (IMs) at a given site conditional on a set of physical parameters (Douglas, 2003). These models form the critical bridge between seismological source, path, and site characterisations and engineering structural analysis and are the key to performing deterministic, probabilistic, and site-specific seismic hazard assessments.

The GMMs over the past several decades have been ones of ever-increasing complexity and refinement. The field has evolved dramatically from simple, empirically derived attenuation laws to a vast and intricate ecosystem of over 500 distinct models, if we account for Openquake's models (Silva et al., 2020). This increase is not only quantitative; it signifies a profound diversity in the approaches used and a more thorough examination of the fundamental physical processes.

Conventional empirical regression, utilising globally compiled strong-motion datasets from significant projects such as NGA-West2 and RESORCE, continues to be a fundamental component (Abrahamson et al., 2014), (Boore et al., 2014). However, the GMMs library has been augmented by a variety of additional methodologies. These include data-driven approaches as Artificial Neural Networks (ANN), which can complex, non-linear relationships in data without a functional form a priori (Derras et al., 2014), and physics-based stochastic simulations, which provide a vital tool for estimating ground motion in data-scarce regions (Atkinson & Boore, 2006). The modern GMMs collectively account for over 42 distinct physical parameters that characterise the earthquake source (e.g., moment magnitude, hypocentral depth, faulting style, stress drop parameter $\Delta\sigma$), the wave propagation path (e.g., various distance metrics, regional quality factors Q, anelastic attenuation, volcanic path effects), and the local site conditions (e.g., time-averaged shear-wave velocity $V_{s30}$, basin depth metrics, high-frequency attenuation kappa, geological unit, and parameters for liquefaction susceptibility).

This development in modelling capability is mirrored, also, by an expansion in the diversity of IMs required for a comprehensive seismic-hazard analysis. The demands of modern engineering have moved beyond the conventional of PGA, PGV, and SA. For instance, a next-generation performance-based seismic design now necessitates a broad suite of IMs. This includes Intensity scales such as Modified Mercalli Intensity (MMI) and Japan Meteorological Agency (JMA) for calibrating models against historical and observational data (Allen et al., 2012); The Modified Mercalli Intensity (MMI) and Japan Meteorological Agency (JMA) scale serve as macroseismic intensity measures which scientists use to validate their models through comparison with historical earthquake data and direct observational records (Allen et al., 2012). The Arias Intensity (IA) and significant duration (RSD) provide better evaluation of total energy input and shaking duration, which makes them suitable for assessing structural damage accumulation and liquefaction potential (Travasarou et al., 2003). The Fourier Amplitude Spectrum (FAS) and Effective Amplitude Spectrum (EAS) frequency-domain representations serve as essential tools for advanced site-response analysis and stochastic ground-motion simulations (Bayless & Abrahamson, 2019). Inelastic spectral displacement (SDi) for structural response, lateral spread displacement (LSD) for ground failure, and period-averaged spectral acceleration (AvgSA) for multi-period demand prediction are few of the parameters that the engineering community has developed to measure specific performance elements (Aristeidou et al., 2023); (Youd et al., 2002); (Weatherill, 2024). (Gulerce & Abrahamson, 2011) state that the vertical motion intensity metrics of vertical spectral acceleration (VHR_SA) and vertical-to-horizontal spectral ratio (V/H) are crucial instruments for the design and assessment of vertically sensitive structures close to fault zones.

This confluence of complexity—diverse GMM methodologies, a vast physical parameter, and a specialized suite of IMs—represents a profound scientific achievement. However, it has created a widening accessibility gap between the frontiers of ground motion science and its practical application by the very community it is meant to serve. The very tools designed for precision have become difficult to wield. The practical application of this state-of-the-art science is often constrained by a significant computational barrier. The use of powerful, open-source engines like the OpenQuake engine is an excellence of transparency and reproducibility (Pagani et al., 2014), but it requires a specialist software environment and considerable programming skills, hence being a deterrent for the majority of practitioners. This overhead also adds to the intrinsic scientific problem of GMM selection.

To bridge this gap between scientific advancement and practical application, this paper introduces OpenGSIM. The objective of this study is to present an automated and conversational workflow that democratizes access to the entire ecosystem of modern ground motion modeling. We detail the development of a novel platform that leverages the ubiquitous Telegram messaging application and the n8n visual automation framework to provide mobile, on-demand access to the complete OpenQuake GMM library. We demonstrate how this tool abstracts the computational and informational complexities, enabling any user to perform seismic prediction by simply responding to a series of guided prompts. A key innovation of the currently approach is the integration of a Large Language Model (Gemini) to provide an automated, AI-powered interpretation of the numerical results, making the outputs more accessible and immediately useful. Ultimately, this work presents a new paradigm for scientific software interaction, transforming a complex analytical task into a simple, conversational, and universally accessible process.

To achieve objectives of this study, the paper will be organised as follows. We begin by creating the full scientific basis, offering a thorough review of the ground motion modeling ecosystem, including its numerous model taxonomies, the extensive physical parameter, and the categorisation of essential Intensity Measures. With this foundation in place, we next go into the design and methodology of the OpenGSIM itself, outlining the function and interaction of its four core layers: Interface, Orchestration, Computation, and LLM. To show how OpenGSIM works, we go through an end-to-end case study, exhibiting the user's conversational journey from first question to final report. Finally, we step back to address the larger implications of our study, evaluating its present limits and possible effect, before making closing comments and proposing intriguing avenues for future research.

## 2 Ground Motion Modeling

Before presenting the automated workflow, it is essential to first understand the complexity and diversity of GMMs. These models are characterised by a tripartite diversity: the theoretical foundations of functional forms, the physical parameters they require, and the broad suite of intensity measures they are designed to predict.

### 2.1 A Taxonomy of Ground Motion Models

GMMs, as implemented in frameworks like the OpenQuake engine, are not a monolithic collection of equations but rather a mosaic of models generated from different techniques and algorithms. While not complete, the important groupings include:

<u>Empirical Models:</u> This is the most typical and frequently utilised method. These models are built by statistical regression on massive strong-motion datasets, fitting a predefined functional form to observed ground motion data. Landmark databases such as NGA-West2 and RESORCE have generated highly complicated empirical GMMs that are now the standard of practice in many fields e.g., (Abrahamson et al., 2014); (Boore et al., 2014); (Akkar et al., 2014).

<u>Stochastic Models:</u> In regions with limited strong-motion recordings, physics-based stochastic ground-motion models are commonly employed. These models, often formulated within the framework of Random Vibration Theory (RVT), generate synthetic ground-motion time series from simplified seismological source representations (e.g., the *Brune $\omega^2$* spectrum (Brune, 1970)) and path attenuation parameters. Such approaches constitute a cornerstone of probabilistic seismic hazard assessment (PSHA) in stable continental regions, where empirical data are sparse (Hanks & McGuire, 1981); (Boore, 2003); (Atkinson & Silva, 2000); (Edwards & Fäh, 2013).

<u>Data-driven models:</u> involves the use of machine learning. These models learn the complex, non-linear relationships between predictor variables and ground motion intensity directly from the data, without imposing functional form a priory. This can provide greater flexibility in capturing observed trends e.g.,(Chaibeddra Tani & Derras, 2024).

<u>Hybrid Models:</u> This category combines elements from the other approaches. For instance, a hybrid model might use simulated ground motions from a physics-based method (like Green's Functions) to supplement empirical data in ranges where recordings are scarce, such as for large magnitudes at short distances (e.g., (Somerville et al., 2001)).

OpenGSIM tool is designed to be agnostic to these methodologies, treating each GMM as a functional object and thereby allowing for the comparison of models with different theoretical underpinnings.

## 2.2    The Physical Parameter

The evolution of GMMs is reflected, also, by the rise of number of the physical parameter they utilize. To perform a calculation, our workflow must be capable of handling over 42 distinct physical parameters, which can be broadly classified, mainly, into three categories (the list of physical all parameters required by GMMs is represented in the Supplementary Material : **Erreur ! Source du renvoi introuvable.**):

<u>Source Parameters:</u> characterise the seismological features of the earthquake rupture.  They generally contain the moment magnitude (Mw), which quantifies the released seismic energy; the hypocentral depth (hypo_depth), controlling attenuation and frequency content; and the faulting mechanism (e.g., rake, dip, strike), which influences radiation pattern and directivity effects. Additional source parameters, such as the stress drop ($\Delta\sigma$), give insights into source dynamics and spectrum scaling, affecting the high-frequency content of ground movements (Brune, 1970); (Atkinson & Silva, 2000).

<u>Path Parameters:</u> define the propagation effects that seismic waves encounter as they travel from the source to the site. These characteristics influence geometric spreading, anelastic attenuation, and crustal heterogeneity along the travel route. Common usual metrics include the Joyner-Boore distance ($R_{JB}$, the closest distance to the surface projection of the rupture) and the rupture distance ($R_{rup}$, the shortest distance to the fault plane). In certain models, the azimuth (Supplementary Material : Table S1)—the angle of the site relative to the fault strike, measured clockwise from North—is also used to account for directivity and hanging-wall effects (Campbell & Bozorgnia, 2008); (Chiou & Youngs, 2014).

<u>Site Parameters:</u> Describe the geological and geotechnical proxies that affect ground-motion amplitude, frequency, and duration. The time-averaged shear-wave velocity in the top 30 metres ($V_{S30}$) is the most often used, it's a proxy for near-surface stiffness and site amplification potential. Current GMMs include depth metrics ($Z_{1.0}$, depth to the 1000 m/s shear-wave velocity horizon; $Z_{2.5}$, depth to 2.5 km/s), high-frequency attenuation ($\kappa_0$), and geological or site-class indicators (e.g., NEHRP or EC8 categories). These factors jointly explain the impedance contrasts and resonance events that regulate local amplification (Seyhan & Stewart, 2014) (Zaoui et al., 2025)(Derras et al., 2020).

## 2.3 A Classification of Modern Intensity Measures

The final dimension of complexity lies in the variety of IMs that an engineer or seismologist may need to calculate. Our workflow provides access to the full suite of IMs available in the OpenQuake engine, which can be grouped by their physical meaning and application (the list of IMs is represented in the Supplementary Material : **Erreur ! Source du renvoi introuvable.**):

<u>Peak Amplitudes:</u> The most conventional IMs, including Peak Ground Acceleration (PGA), Velocity (PGV), and Displacement (PGD). They are commonly utilised in design codes and seismic hazard assessment.

<u>Spectral Ordinates:</u> This large family describes the response of a single-degree-of-freedom oscillator. It includes Pseudo-Spectral Acceleration (SA) at various periods, as well as the recently developed Average Spectral Acceleration (AvgSA), which provides a robust measure of structural demand by averaging spectral ordinates across a range of periods.

<u>Energy and Duration Measures:</u> These IMs, which include Arias Intensity (IA), Cumulative Absolute Velocity (CAV), and several definitions of significant duration (such as RSD575 and DRVT), measure the energy content and duration of shaking. This makes them more reliable predictors of the potential for cumulative damage than just peak values.

<u>Frequency-Domain Measures:</u> The Fourier Amplitude Spectrum (FAS) and the Effective Amplitude Spectrum (EAS) provide a direct view of the frequency content of the ground motion and are essential for stochastic simulations.

<u>Engineering and Damage-Proxy Measures:</u> This category includes IMs designed to predict specific types of damage, such as Inelastic Spectral Displacement (SDi) for performance-based design and Lateral Spread Displacement (LSD) for liquefaction hazard assessment.

<u>Intensity Scales and Ratios:</u> This includes observational scales like Modified Mercalli Intensity (MMI) and the Japan Meteorological Agency (JMA) scale, as well as the Vertical-to-Horizontal Spectral Ratio (VHR_SA), which is particularly informative in near-fault and site characterization contexts.

# 3  System Architecture and Methodology

To address the challenges outlined in the previous sections (assembling all GMMs in the same ecosystem), we designed and implemented a Telegram (Telegram, 2025) chatbot (named OpenGSIM) through an application programming interface. The n8n platform (n8n, 2025), a conversational workflow for on-demand GMM calculation, fully automates OpenGSIM. The resulting architecture is a microservices-based system orchestrated by a visual workflow automation platform.

## 3.1  Conceptual Overview

The overall architecture of the OpenGSIM is illustrated in Figure 1. The system operates as a stateful, event-driven application where the user's messages on Telegram act as triggers that propel the workflow through a series of defined states. The n8n automation platform serves as the central orchestration engine, receiving user inputs via a webhook (we used ngrok (ngrok, 2025) proxy service) and dispatching tasks to the appropriate backend services. The scientific calculations are delegated to a dedicated (Python Software Foundation, 2025) wrapper that interfaces with the OpenQuake engine, running within an isolated Docker container (Merkel, 2014). The numerical results are then post-processed and, optionally, sent to the Gemini API for natural-language (Google, 2025) interpretation before the final report through (WeasyPrint, 2025) and data files are delivered back to the user via the Telegram Bot API.

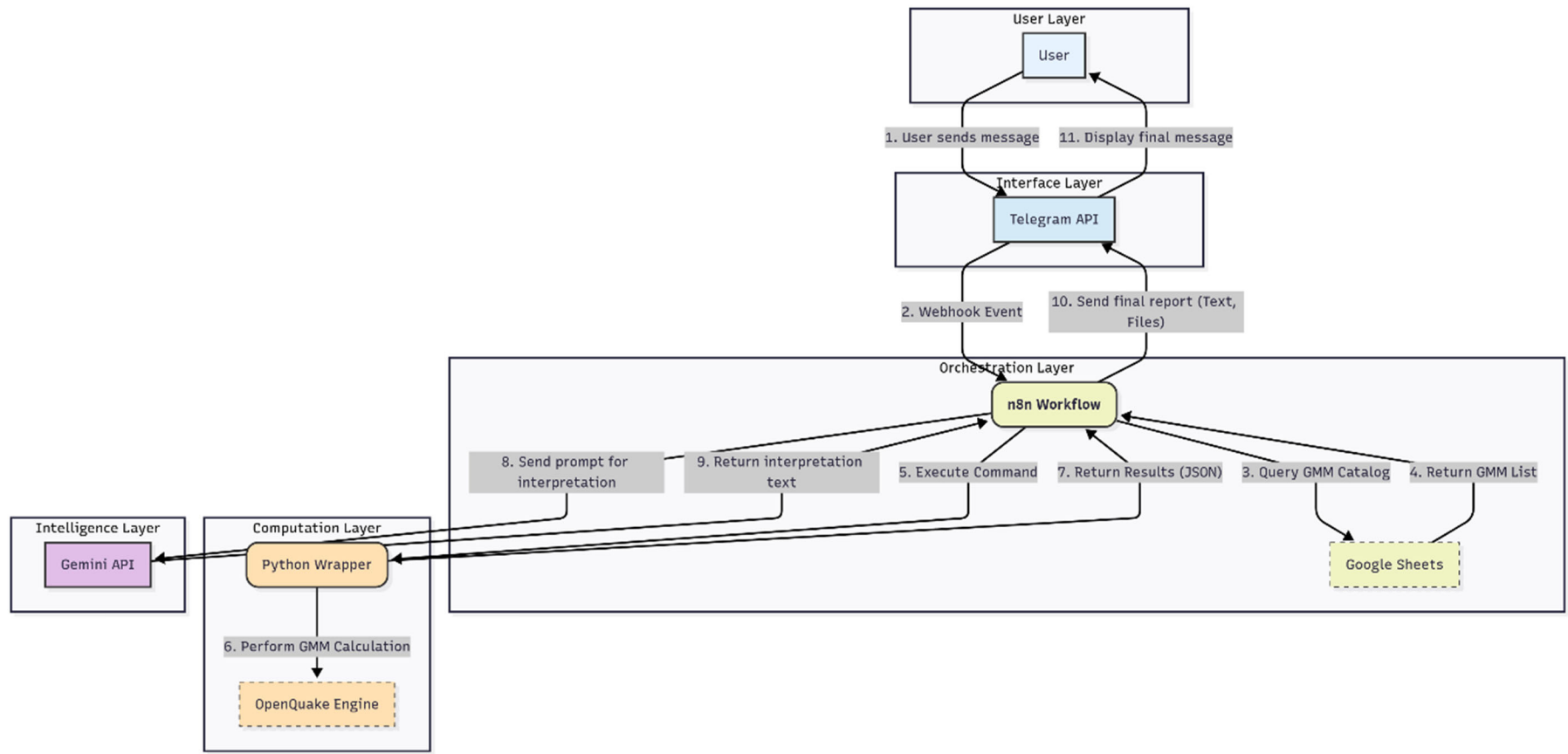

Figure 1 : Conceptual architecture of the OpenGSIM. The user interacts with the Telegram Bot. Messages are sent via a webhook to the n8n workflow, which acts as the central orchestrator. n8n queries a Google Sheet to retrieve the GMM catalogue. Based on user choices, n8n executes a Python script within a dedicated Docker container, which leverages the OpenQuake-engine for the GMM calculation. The results sent to the Gemini API for interpretation. Finally, n8n formats and delivers the final report and data files back to the user via Telegram.

## 3.2 The User and Interface Layers

The top layer of the architecture is dedicated to user interaction (Figure 1). The User Layer represents the end-user (e.g., an engineer, researcher, or student) interacting with their device. The Interface Layer is handled exclusively by the Telegram Bot API. This choice was strategic: Today, Telegram reportedly has more than 1 billion monthly active users. Ranked by the number of downloads, Telegram is among the top 10 most popular social networks in the world (Telegram, 2025). Through Telegram, the user selects for n8n via the ngrok webhook the type of IMs, the GMM class that gives IM. In addition, he provides the physical parameters supported by the GMM class.

This layer is the route for all user interactions (see Figure 1, Arrows 1 and 11). Our process uses webhook capability, such as (ngrok, 2025)which sends any user communication as a JSON-formatted payload to a pre-registered HTTPS endpoint. This event-driven method enables the orchestration Layer to respond to user input in real time.

## 3.3 The Orchestration Layer

The central system of OpenGSIM is the Orchestration Layer (Figure 1)(ngrok, 2025), managed by an n8n workflow. n8n is an open-source, node-based automation tool that allows for the visual construction of complex processes, making the bot's logic transparent, debuggable, and easily modifiable. This layer receives events from the Interface Layer (Arrow 2) and coordinates all backend tasks.

The main workflow functions as a state machine guiding the user through the calculation:

Session Management and Data Retrieval: Upon receiving a user's choice, Code nodes query a Google Sheet (Arrows 3 & 4), which acts as a simple, easily maintainable database for the 500+ GMMs catalogue and their associated metadata.

Logic and Validation: If nodes and code nodes are used to validate user input, ensuring the correct number and format of physical parameters are provided before proceeding (see Figure 3 and Figure 5).

Task Delegation: The orchestrator dispatches tasks to the appropriate layers. It invokes the Computation Layer by assembling and running a command-line instruction via an Execute Command node (Arrow 5). It then sends the numerical results to the Intelligence Layer for analysis via an HTTP Request node (Arrow 8).

Response combination and delivery: Finally, the workflow gathers all the artefacts—the raw JSON from the computation layer (Arrow 7), the text from the intelligence layer (Arrow 9), and any generated files—and delivers the final prepared report back to the user through the interface layer (Arrow 10).

## 3.4 The Computation Layer

This layer is the scientific core of the system, responsible for performing the GMM calculations and IMs prediction. It is encapsulated within an isolated Docker container to ensure a consistent and reproducible computational environment. This layer consists of two main components:

The Python Wrapper: a custom script, providing a command-line-driven interface to the OpenQuake library. It is built for flexibility, dynamically importing the GMM class supplied by the user and populating the OpenQuake context objects (Sites context : site effect parameters, Rupture context : source effect parametrs, Distances context: path effect parameters) : these parameters (proxies) are provided par user (e.g. Figure 5). The script intelligently finds all IMs allowed by the chosen model, conducts the computations, and writes the aggregated results as a single JSON object to output, allowing for simple reading by n8n.

The OpenQuake Engine: the underlying scientific library (Pagani et al., 2014); (Silva et al., 2020) that performs the validated GMM calculations (Arrow 6). Our wrapper directly utilises its "get_mean_and_stddevs" function to ensure results are consistent with the standard of practice.

## 3.5 The LLM (Large Language Model) Layer

To bridge the gap between raw numerical output and actionable insight, we implemented an Intelligence Layer powered by a Large Language Model. After a successful computation, the Orchestration Layer sends the key spectral results to this layer (Arrow 8).

For this task, we interface with the Google Generative Language API to utilise the Gemini 1.5 Flash model (it is a free LLM). A carefully designed prompt instructs the model to act as a seismological expert, analyse the provided spectral data (e.g., SA or EAS values versus period/frequency), and generate a brief, qualitative summary. This interpretation typically includes identifying the peak spectral response and commenting on its potential implications for structures. The choice of the "Flash" model was motivated by its excellent balance of analytical capability, low latency, and cost-efficiency, making it highly suitable for a real-time interactive application. The resulting text is then sent back to the orchestrator for inclusion in the final report (Arrow 9).

## 4   The OpenGSIM Interface and User Workflow

The architecture described above is presented to the end-user through a conversational interface within the Telegram application. The entire process is intuitive, guiding the user through a series of simple, numbered steps without requiring any prior knowledge of the underlying system (n8n and OpenQuake).

Upon commencing a compute session with the "/start" command, the user is met with a welcome message: the bot's description section presents an overview of its capabilities ("What can this bot do?"), alerting the user that they may access over 500+ GMMs and 21 IMs. Following, the bot displays the compact user manual, as seen in Figure 2. This opening message covers the four key phases of the workflow: (1) IM selection, (2) GMM class selection, (3) physical parameters (inputs), and (4) reception of results.

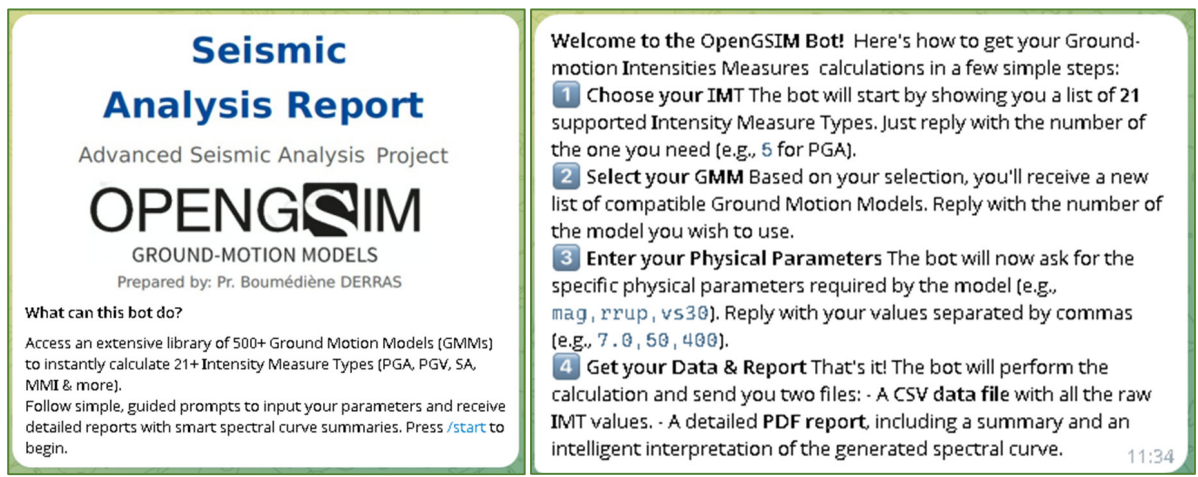

Figure 2 : The initial user interface of the OpenGSIM. On the left, the welcome message triggered by the /start command. On the right, the bot's static profile description summarizing its capabilities, which outlines the four-step procedure for performing a calculation.

This combination of an introductory instruction and a static explanation guarantees that the user is fully aware of the bot's goal and operational process from the very beginning. The process progresses via a basic reply-based interaction model: the bot provides a numbered list of alternatives, and the user answers with the matching number to make their pick. This approach was chosen for its universality and simplicity, limiting the chance for mistakes. Therefore, the first interactive step of the workflow is the IM selection. As illustrated in the Supplementary Material (**Erreur ! Source du renvoi introuvable.**), the bot presents a numbered list of the IMs available, each accompanied by a brief scientific definition. Once the user replies with the number corresponding to the IM chosen, the orchestration layer (Figure 1) uses this choice to perform the critical task of filtering the entire GMM catalogue. In this context, an example is represented in Figure 3.

Now that the overall architecture and user workflow have been established, the following section will demonstrate this entire process in action through a detailed case study.

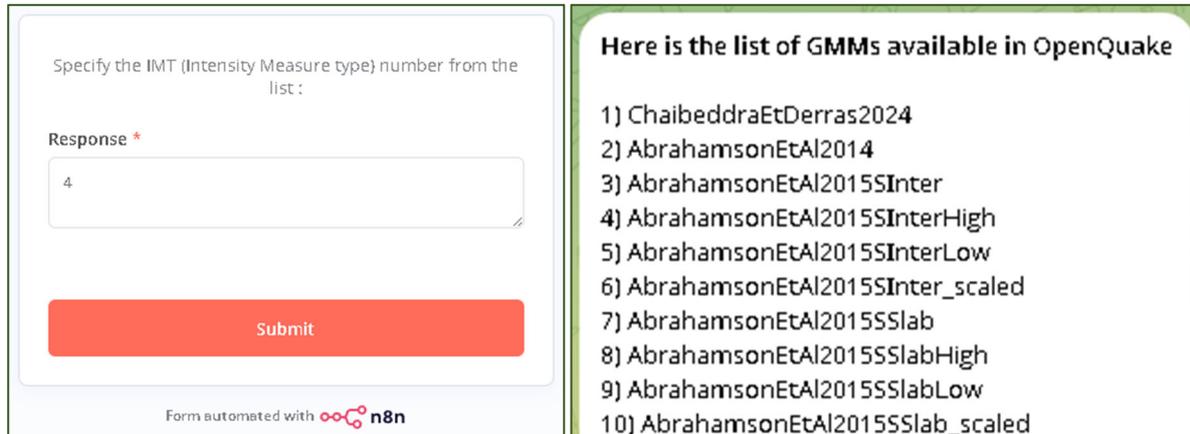

figure 3 : The two-step selection process within the OpenGSIM workflow. On the left, the user submits their choice of IM (in this case, "4", corresponding to SA Spectral Acceleration; Supplementary Material **Erreur ! Source du renvoi introuvable.**). On the right, the bot responds with a filtered and numbered list of the 500+ GMMs that support the selected IM.

## 5    Results: A Case Study Workflow

To show the end-to-end capabilities of the OpenGSIM, we give an example of entire computation workflow for a practical seismic scenario. This case study explains the method of determining the 5%-damped pseudo-spectral acceleration (SA) using the known AbrahamsonEtAl2014 (ASK14) GMM in their basic version.

### 5.1    Scenario Definition and User Input

After to have selected the IM's number; the user chooses the number of GMM class that he would like used. For instance, the number 2 is corresponded of the AbrahamsonEtAl2014 (see the left panel of Figure 3).

After choosing the AbrahamsonEtAl2014 model, the workflow shows the user a short summary of the chosen GMM, which includes a short description and a list of all the physical parameters needed, organised by category (Rupture, Distance, Site). The bot also goes over each required parameter one by one. It gives a clear scientific definition of each parameter, the unit it should be in, and a typical range of values. This contextual guidance is very important for reducing error and also serves an educational purpose by making sure the user knows what the values they are giving means (Figure 4).

[Figure 4 image: Two chat panels. Left panel describes the ASK14 GMPE and lists Rupture: dip,mag,rake,width,ztor; Distance: rjb,rrup,rx,ry0; Site: vs30,vs30measured,z1pt0. Physical Parameters: vs30 definition, Unit m/s, Typical Range 150 (soft soil) to 2000 (rock). Right panel lists Intensity Measure Types: PGA, PGV, SA with definitions, units, and typical ranges.]

Figure 4 : The contextual definitions phase. In the right panel the bot first provides a description of the selected GMM (ASK14) and its required inputs. It then details each parameter (e.g., vs30) with its definition, unit, and typical range before prompting the user for all values. The right panel shows a similar informational card for the supported IMs.

Subsequently, the workflow prompts for the 12 physical parameters required by the ASK14 model. For this scenario, we defined a crustal earthquake and provided the corresponding values in the requested comma-separated format, as shown in Figure 5. The input physical parameter values for this case study are dip = 70° (fault dip angle), mag = 6.5 (moment magnitude), rake = 90° (fault rake angle), width = 10 km (fault width), ztor = 2.0 km (depth to top of rupture), rjb = 20 km (Joyner-Boore distance), rrup = 25 km (closest distance to rupture), rx = 15 km (horizontal distance), ry0 = 0 km (site coordinate), vs30 = 500 m/s (shear-wave velocity), vs30measured = 1 (and 0 for inferred Vs30), and z1pt0 = 50 m (depth for a vs = 1000 m/s).

[Figure: Form prompting "Please enter the physical parameter values according to the following format: dip,mag,rake,width,ztor,rjb,rrup,rx,ry0,vs30,vs30measured,z1pt0" with input "70,6.5,90,10,2,20,25,15,0,500,1,50" and a Submit button. Form automated with n8n.]

Figure 5 : The parameter elicitation step for the case study. The bot requests the specific inputs for the ASK14 model, and the user replies with the corresponding 12 values in a single, comma-separated message.

## 5.2 Output: Report, Data, and AI-Powered Interpretation

The Orchestration Layer starts the Computation Layer (Figure 1) after the input parameters have been successfully checked. The system does the math for all supported spectral periods and sends the results to the user within seconds (Figure 6, Figure 7, Figure 8). The final deliverables for this case study include:

A Visual Summary: A plot of the response spectrum is generated and sent as a PNG image, providing an immediate visualisation of the mean spectral acceleration and the ±s (standard deviation) range across all periods (Figure 6).

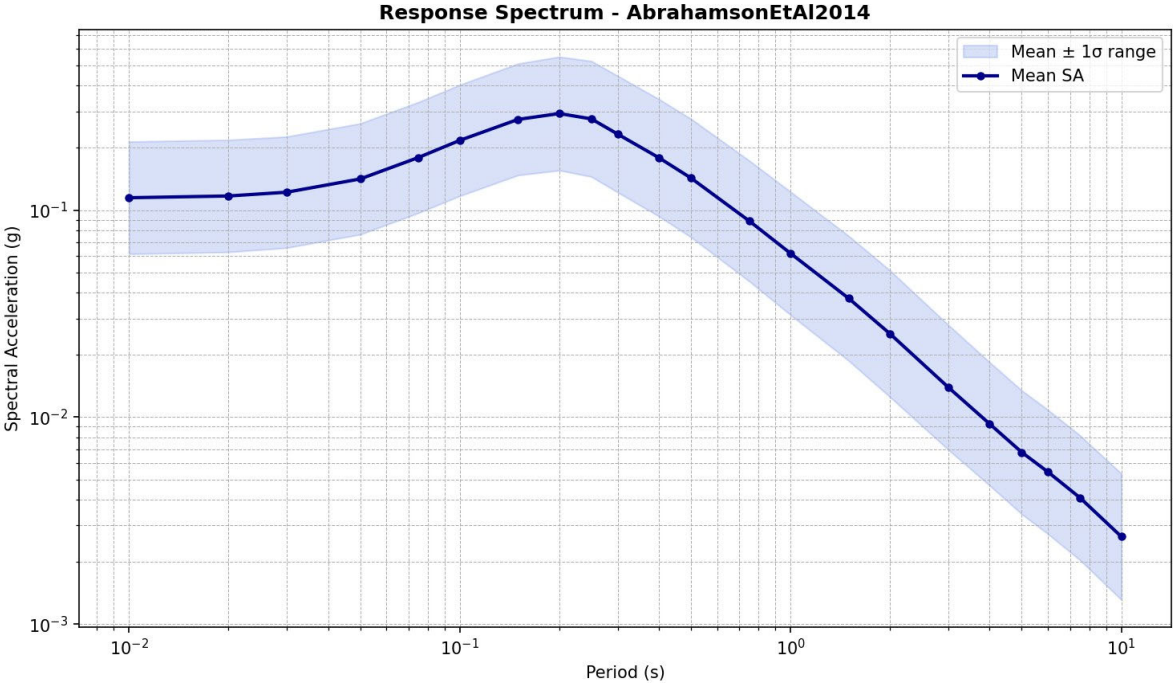

Figure 6 : The generated response spectrum plot, visualizing the mean SA and its aleatory variability for ASK14 model.

A Textual Report: A Textual Report: A formatted text message summarises the physical parameter values (inputs) and presents a sampled summary of the key spectral values (see Figure 7).

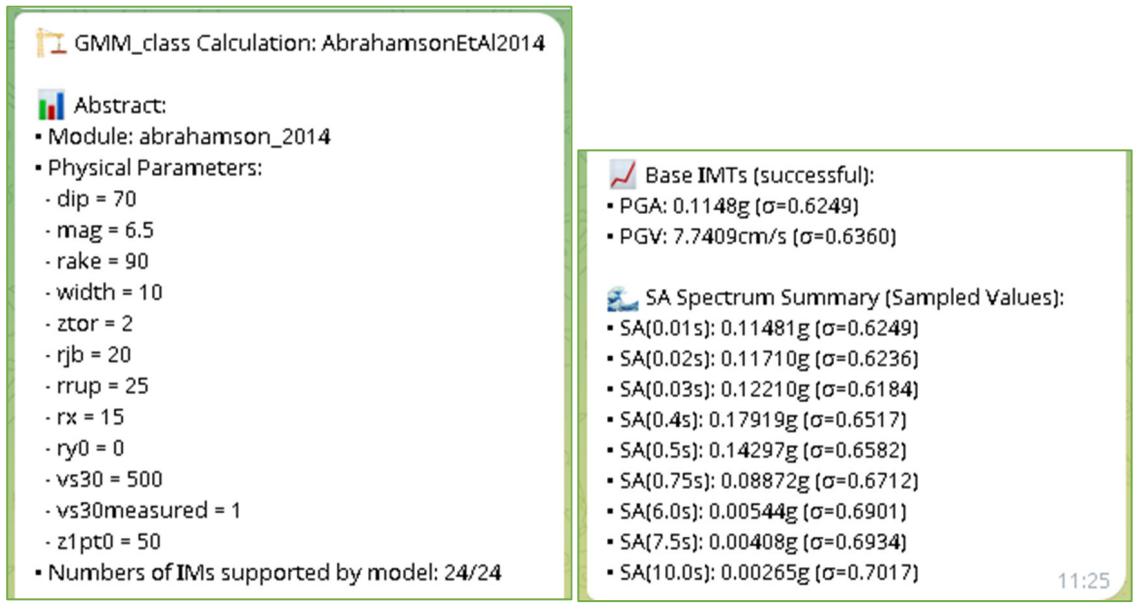

Figure 7: the summary report, delivered in-chat, recalling the input and output values of the ASK14 model.

**Reporting and AI-Powered Interpretation**: Upon successful computation, the workflow triggers the intelligence layer (see Figure 1) and subsequently delivers the final results to the user. This output includes a downloadable CSV file with the raw numerical data and a comprehensive PDF report. A key innovation of the OpenGSIM is the inclusion of an automated analysis within this report, generated by the intelligence layer. The numerical spectral results are passed to the Gemini 1.5 Flash model through a structured prompt, which instructs the AI to act as an earthquake engineering expert and provide a qualitative interpretation. The design of this prompt is critical to obtaining a consistent and relevant analysis. The specific prompt used in this workflow is detailed in Table 1.

Table 1 : The structured prompt sent to the Gemini 1.5 Flash API for the interpretation of Spectral Acceleration (SA) data.

| Role: | you are an earthquake engineering expert. your task is to interpret spectral acceleration (SA) data. |
|---|---|
| Context: | The data provided is acceleration in units of 'g'. |
| Task: | Based strictly on the data provided, provide an analysis following these exact steps. Use the term "spectral acceleration" in your response: |
| | Overall Summary: One-sentence summary of the seismic hazard level. |
| | Peak Acceleration: State the peak spectral acceleration and the period at which it occurs. |
| | Implications for Short-Period Structures: Describe the implications for rigid buildings (e.g., low-rise masonry). |

| | Implications for Long-Period Structures: Describe the implications for flexible buildings (e.g., high-rise towers). |
|---|---|

The Gemini API provided the interpretation shown in Figure 8 for our case study using the ASK14 model. The model of study appropriately recognises the PGA at a short period (0.2 s) and compares the seismic demand on rigid, short-period buildings to the reduced demand on flexible, long-period buildings for this study scenario. It is important to note that although this automatic interpretation provides a helpful summary, it is meant to assist the analyst as an initial analytical evaluation and is not a substitute for a full evaluation by a certified engineering expert.

Finally, this case study demonstrates the successful implementation of our design philosophy, showcasing the system's ability to abstract a complex GMM calculation into an efficient, guided, and conversational workflow. The process culminates in a rich, multi-format, and intelligently interpreted output, transforming a traditionally laborious task into an accessible, on-demand analysis.

> **AI-Powered Interpretation**
>
> Here is the interpretation of the **Spectral Acceleration (SA)** data:
>
> 1. Overall Summary: This spectral acceleration data suggests moderate to high seismic hazard potential, particularly for structures with shorter periods.
>
> Page 2 of 7
>
> 2. Peak Acceleration: The peak spectral acceleration is 0.2929 g, occurring at a period of 0.2 seconds.
>
> 3. Short-Period Structures: The relatively high spectral acceleration values at short periods (e.g., 0.01s to 0.3s) indicate a significant seismic demand on stiff or rigid structures, implying these structures would experience higher accelerations.
>
> 4. Long-Period Structures: The spectral acceleration values decrease significantly at longer periods (e.g., 1.0s to 10.0s), indicating a lower seismic demand on flexible structures compared to rigid ones, but the accelerations are still non-negligible and must be considered.
>
> This interpretation was generated by a Gemini 2.0 Flash AI agent and should be reviewed by a qualified expert.

Figure 8 : An example of the AI-powered interpretation generated by the Gemini 1.5 Flash model for the ASK14 case study results (pdf report). The output follows the structured four-point format defined in the prompt

## 6   Discussion

The OpenGSIM Telegram bot, as demonstrated, represents more than just a technical tool; it offers a new practical paradigm of earthquake engineering. By capturing the complexity and the diversity of the Ground-Motion Models (GMMs) within an interactive interface. This study aims to address the critical accessibility gap that has arisen between cutting-edge GMMs and the broader community of practitioners. Without a dedicated framework like OpenGSIM, practitioners are typically faced with two challenging options: either manually reimplementing a GMM's functional form from its source publication, a process prone to error, or installing and configuring a complex computational environment such as the OpenQuake engine. OpenGSIM is a platform-independent tool that eliminates the traditional barriers of software installation and complex configuration, making advanced seismic analysis accessible on any device and under any OS (operating system). In essence, this work takes ground motion prediction—once the exclusive domain of specialists—and places it directly into the hands of researchers, engineers, technicians, and students. The bot flattens this steep learning curve, allowing it to be possible in a related field to perform a sophisticated, scenario-based computation in minutes from a mobile device. This has profound implications for both practice and education.

This BOT offers a robust capability for civil engineers, allowing them to swiftly perform computation of IMs. In addition, the BOT simplifies the process of comparing several GMMs for a given seismic scenario, an essential feature for projects focused on specific projects where design spectra based on codes may not be applicable. This capability for rapid GMM comparison is crucial for quantifying epistemic uncertainty—the variability in predictions due to limited scientific knowledge—which is a fundamental consideration for any robust and reliable engineering design. This BOT allows students and educators to transform GMMs from abstract equations into interactive tools, fostering a more intuitive understanding of how physical parameters like magnitude, distance, and local site conditions (Vs30, z1.0) influence the resulting ground motion.

The technical innovation of this work is using n8n workflow and the integration of a Large Language Model (LLM) for automated interpretation of spectral curves. The Gemini 1.5 Flash, used here, correctly identifies the spectral variations, such as the period of peak spectral response, and translates this numerical information into meaningful engineering information. However, the LLM is not a substitute for the judgement of a qualified engineering expert who can account for site-specific non-linearities, structural detailing, and the full context of a design project. The role of the LLM here is that of an assistant for initial analysis, while the final engineering decision remains firmly in the domain of the human expert.

# 7 Conclusion

This paper introduces the OpenGSIM Telegram bot, a new way to link the complexities of today's ground motion studies with the practical needs of those who work in earthquake engineering and seismology. The research has demonstrated that it is possible to make advanced seismic analysis tools very easy to use without sacrificing scientific accuracy by putting a large group of over 500 classes of GMMs, a complete set of 42 physical parameters and 21 intensity measures, and an AI-powered understanding layer into a simple, widely available Telegram interface.

Instead of a new GMM, this product provides a new way to use a large number of GMMs. This tool greatly speeds up the study of uncertainty for the seismic hazards assessment by making it easy to compare models with different setups and basic ideas, from simple empirical equations to random simulations. The bot gives design spectra for specific situations and a group of advanced IMs (like SDi or AvgSA) that are immediately useful for performance-based design for a structural engineer, allowing for a better understanding of how structures act, going beyond just peak acceleration. The ability to quickly calculate IMs like LSD or Arias Intensity for a specific situation provides a useful first step for checking the risk of liquefaction for a geotechnical engineer.

The addition of an AI understanding layer, even though it is new, points to a future where scientific software not only creates data but also begins the process of developing scientific knowledge. By automatically doing the first analysis of a spectral curve, the bot allows the expert to focus on more complex evaluation and making decisions. Future plans will focus on increasing the system's abilities to include processing many things at once and analysing records customised for each user.

In the end, we believe that this conversational model represents a promising direction for scientific software, making strong computing tools easier, more accessible, and more important for the global group dedicated to reducing seismic risk.

# 8 Acknowledgments

The author wants to say thank you for the strong, free tools and systems that were very helpful in making the OpenGSIM. It would not have happened without the OpenQuake-engine codes, created by the Global Earthquake Model (GEM) Foundation, which is the main science part of what we made. We want to thank the people who made and take care of the n8n project for building a tool that lets us automate tasks in a way that is both easy to change and strong. We also want to say that the Telegram Bot API was indispensable for us because it gave us a universal tool that anyone can use, and Python was useful because it has many scientific packages. Making and putting this project out was made much easier because we used Docker to put everything in containers, WeasyPrint to make PDF files, the Google Sheets API which worked as a simple database for the GMM catalogue, and the safe tunnelling from ngrok while we tested it. The writer also wants to thank Google's Gemini Large Language Model for its help with the conception and writing of this paper. We also want to thank the Algerian Directorate General for Scientific Research and Technological Development (DGRST) for their support.## 9 Data and Code Availability

The Python wrapper scripts, the complete n8n workflow (JSON format), and all materials developed for this study will be openly available on GitHub. The OpenGSIM can be accessed and used directly on the Telegram platform, in the first time only, for reviewers.

## 10 References

Abrahamson, N. A., Silva, W. J., & Kamai, R. (2014). Summary of the ASK14 ground motion relation for active crustal regions. *Earthquake Spectra*, *30*(3), 1025–1055. https://doi.org/10.1193/070913EQS198M

Akkar, S., Sandıkkaya, M. A., & Bommer, J. J. (2014). Empirical ground-motion models for point- and extended-source crustal earthquake scenarios in Europe and the Middle East. *Bulletin of Earthquake Engineering*, *12*(1), 359–387. https://doi.org/10.1007/s10518-013-9461-4

Allen, T. I., Wald, D. J., & Worden, C. B. (2012). Intensity attenuation in active crustal regions. *Journal of Seismology*, *16*(2), 409–433.

Aristeidou, S., Tarbali, K., & O'Reilly, G. J. (2023). A ground motion model for orientation-independent inelastic spectral displacements from shallow crustal earthquakes. *Earthquake Spectra*, *39*(3), 1601–1624.

Atkinson, G. M., & Boore, D. M. (2006). Earthquake ground-motion prediction equations for eastern North America. *Bulletin of the Seismological Society of America*, *96*(6), 2181–2205. https://doi.org/10.1785/0120050245

Atkinson, G. M., & Silva, W. J. (2000). Stochastic modeling of California ground motions. *Bulletin of the Seismological Society of America*, *90*(2), 255–274. https://doi.org/10.1785/0119990064

Bayless, J., & Abrahamson, N. A. (2019). Summary of the BA18 ground-motion model for Fourier amplitude spectra for crustal earthquakes in California. *Bulletin of the Seismological Society of America*, *109*(5), 2088–2105.19

# Supplementary Material

Boumédiène DERRAS[1,2]

[1]Civil engineering & hydraulic department, university of Dr Moulay Tahar, Saida, Algeria. boumediene.derras@univ-saida.dz. orcid.org/0000-0002-3979-7083

[2]Risk Management and assessment (RISAM) laboratory, University of Abou Bekr Belkaid, Tlemcen, Algeria, boumediene.derras@univ-tlemcen.dz

Table S1 : Description of all physical parameters required by GMMs in OpenGSIM

| Effects | Name | Definition | Unit |
|---|---|---|---|
| Source effect | $\Delta\sigma$ | Stress Drop. A parameter describing the stress reduction on the fault during an earthquake. Used in some advanced GMMs. | bar |
| | mag | Moment Magnitude (Mw). A logarithmic scale measuring the total energy released by an earthquake. | none |
| | Hypo depth | Depth of the earthquake's hypocenter (point of initiation) below the surface. | km |
| | hypo_lat | Latitude of the hypocenter. | degrees |
| | hypo_lon | Longitude of the hypocenter. | degrees |
| | dip | The angle of the fault plane with respect to the horizontal. | degrees |
| | rake | The angle describing the direction of slip on the fault plane. Defines the style of faulting. | degrees |
| | ztor | Depth to the Top of Rupture. The shallowest point on the fault rupture plane. | km |
| | width | The down-dip width of the fault rupture plane. | km |
| | in_cshm | in_cshm: A Boolean flag (1=true, 0=false) indicating if the earthquake source is within the Canterbury zone. Canterbury Seismic Hazard Model (CSHM) to apply region-specific adjustments for New Zealand (Bradley, 2013). | boolean (1/0) |
| Path effect | rrup | Rupture Distance. The shortest distance from the site to any point on the fault rupture plane. | km |
| | rjb | Joyner-Boore Distance. The shortest horizontal distance from the site to the surface projection of the rupture plane. | km |
| | repi | Epicentral Distance. The horizontal distance from the site to the earthquake's epicenter. | km |
| | rhypo | Hypocentral Distance. The straight-line distance from the site to the earthquake's hypocenter. | km |
| | rvolc | (Context-dependent, Japan) Source to site distance passing through surface projection of volcanic zone. | km |
| | rx | Horizontal distance from the site to the fault trace, measured perpendicular to the trace. Can be positive or negative. | km |
| | ry0 | Horizontal distance from the site to the fault trace, measured parallel to the trace from the midpoint. | km |
| | azimuth | The angle of the site relative to the fault strike, measured clockwise from North. | degrees |
| | rcdpp | Closest Distance to the Centred Predictor Point. A specific geometric parameter used to model the effects of near-fault forward directivity. A smaller value indicates a higher potential for strong directivity pulses. | km |
| Site effect | vs30 | Time-averaged shear-wave velocity in the top 30 meters of soil. It is a key proxy for site amplification, with lower values indicating softer soil. | m/s |
| | vs30 measured | A Boolean flag (1 for true, 0 for false) indicating if the Vs30 value comes from a direct measurement or an inference. | boolean (1/0) |
| | z1pt0 | Depth to the shear-wave velocity horizon of 1.0 km/s. It characterizes the basin depth. | m |
| | z1pt4 | Depth to the shear-wave velocity horizon of 1.4 km/s. Used in some Japanese models. | m |
| | z2pt5 | Depth to the shear-wave velocity horizon of 2.5 km/s. It characterizes the deeper basin structure. | km |
| | backarc | A Boolean flag (1 for true, 0 for false) indicating if the site is in a back-arc tectonic setting, which affects wave attenuation. | boolean (1/0) |
| | xvf | (Context-dependent, Japan) Horizontal distance from the site to the volcanic front line. It's positive in the fore-arc region. | km |
| | PHV | Amplitude (Peak) of the Horizontal-to-Vertical Spectral Ratio (H/V). Measures the site's primary amplification factor. | none |
| | THV | Fundamental period (Time) of the site, corresponding to the peak of the H/V ratio. | s |
| | slope | Ground slope, used for liquefaction models (LSD). Represents the gradient in percent. | % |
| | Free face ratio | Free-face ratio, used for liquefaction models (LSD). Ratio of the height of a free face to its distance from the site. | none |
| | T_15 | Cumulative thickness of saturated sandy layers with corrected SPT blow-counts (N1)60 < 15. Used for liquefaction models. | m |
| | F_15 | Average fines content for the T_15 layers. Used for liquefaction models. | % |
| | D50_15 | Average mean grain size for the T_15 layers. Used for liquefaction models. | mm |
| | kappa0 | High-frequency filter parameter representing near-surface attenuation directly beneath the site. | s |
| | f0 | Fundamental site frequency, or resonance frequency. | Hz |
| | bas | Basin parameter, typically a boolean flag (1/0) indicating if the site is on a sedimentary basin. | boolean (1/0) |
| | lat | Site latitude. | degrees |
| | lon | Site longitude. | degrees |
| | region | Geologic region string, used for regional adjustments in some GMMs.<br><br>1: Iberian Peninsula Includes Spain and Portugal.<br><br>2: France and Germany Includes France, Germany, Belgium, the Netherlands, Switzerland and the United Kingdom. It is a region of low to moderate seismicity.<br><br>3: Italy and the Western Balkans Includes Italy, Austria, Slovenia, Croatia, Serbia, etc. It is a region with high seismicity. It is a great default choice for our tests because it is seismologically very active.<br><br>4: Greece and Turkey Includes Greece, Albania and the western part of Turkey. It is the most seismically active region in Europe.<br><br>5: Pannonian Basin Mainly includes Hungary and the surrounding regions.<br><br>0 (or unspecified): If you do not specify a region or if you give a value of 0, the model will use the basic coefficients of Kotha et al. (2020), which represent an average for the dataset. | 0 to 6 |
| | Geology | Gelogic age: CENOZOIC, HOLOCENE, JURASSIC-TRIASSIC, CRETACEOUS, PALEOZOIC, PLEISTOCENE, PRECAMBRIAN, UNKNOWN | Name of Gelogic age |

| | siteclass | EC8 site category. | A to E |
|---|---|---|---|
| | | Class A: Rock or other rock-like geological formation, including at most 5 m of weaker material at the surface. (Vs30 > 800 m/s) | |
| | | Class B: Deposits of very dense sand, gravel, or very stiff clay, at least several tens of meters thick. (360 m/s ≤ Vs30 ≤ 800 m/s) | |
| | | Class C: Deep deposits of dense or medium-dense sand, gravel, or stiff clay. (180 m/s ≤ Vs30 ≤ 360 m/s) | |
| | | Class D: Deposits of loose-to-medium cohesionless soil or soft-to-firm cohesive soil. (Vs30 < 180 m/s) | |
| | | Class E: A soil profile consisting of a surface alluvium layer with Vs30 < 360 m/s, thickness ≥ 20 m, overlying stiffer material. | |
| | soiltype | A simplified soil classification, Soiltypes - 1 : Hard rock, 2 : very stiff soil, 3 : Dense soil, 4 : Medium-dense soil, 5 : Soft soil, 6 Very soft-. If soiltype > 6, OpenQuake automatically defaults to 1 (hard rock). | 1 to 6 |

Table S2 : Description of all IMs supported in OpenGSIM bot

| N° | Name | Definition | Unit | The number of GMMs Classes by IM |
|---|---|---|---|---|
| 1 | PGA | Peak Ground Acceleration. The maximum acceleration experienced by the ground during shaking. | g | 472 |
| 2 | PGV | Peak Ground Velocity. The maximum velocity experienced by the ground during shaking. | cm/s | 256 |
| 3 | PGD | Peak Ground Displacement. The maximum displacement of the ground during shaking. | cm | 6 |
| 4 | SA | Spectral Acceleration. The max acceleration of a single-degree-of-freedom oscillator. Requires a period (T) and 5% damping is assumed unless specified. | g | 462 |
| 5 | IA | Arias Intensity. A measure of the total energy of the ground motion over its entire duration. | m/s | 11 |
| 6 | CAV | Cumulative Absolute Velocity. The integral of the absolute value of the acceleration time series, a measure of damaging potential. | g.s | 7 |

| 7 | MMI | Modified Mercalli Intensity. A qualitative scale (I-XII) that describes the severity of earthquake shaking based on observed effects. | integer | 12 |
|---|---|---|---|---|
| 8 | RSD2080 | Significant Duration between 20% and 80% of cumulative IA | s | 2 |
| 9 | RSD575 | Significant Duration between 5% and 75% of cumulative IA | s | 8 |
| 10 | RSD595 | Significant Duration between 5% and 95% of cumulative IA | s | 7 |
| 11 | FAS | Fourier Amplitude Spectrum. A representation of the ground motion's amplitude at different frequencies. Requires a frequency (f) in Hz. | g.s | 2 |
| 12 | SDi | Inelastic Spectral Displacement. The peak displacement of a structure that behaves inelastically. Requires period (T) and ductility or strength ratio (R). | cm | 2 |
| 13 | LSD | Lateral Spread Displacement. The permanent horizontal ground displacement due to soil liquefaction. Typically, from models like (Youd et al., 2002); (Zhang & Zhao, 2005). | m | 4 |
| 14 | AvgSA | Average Spectral Acceleration over a range of periods. The specific period band must be defined in the model. | g | 3 |
| 15 | JMA | Japan Meteorological Agency seismic intensity scale, an instrumental scale ranging from 0 to 7 used in Japan. | integer | 7 |
| 16 | SA1 | Spectral Acceleration. The max acceleration of a single-degree-of-freedom oscillator. At period (T) equal to 1.0 s and 5% damping is assumed unless specified. | g | 2 |
| 17 | EAS | Effective Amplitude Spectrum, derived from the Fourier Amplitude Spectrum (FAS). Represents a smoothed, band-averaged measure of ground motion amplitude in the frequency domain. Useful for site response and stochastic simulations. | cm/s² | 1 |
| 18 | VHR_SA | VHR is the ratio of the 5 %-damped vertical response spectrum (SA) to the geometric mean of the two horizontal components, all expressed in natural-log units in the regression. It quantifies | none | 1 |

| | | how much vertical shaking is expected relative to horizontal shaking for a given oscillator period. | | |
|---|---|---|---|---|
| 19 | VHR_PGA | VHR is the ratio of peak ground acceleration (PGA) to the geometric mean of the two horizontal components, all expressed in natural-log units in the regression. It quantifies how much vertical shaking is expected relative to horizontal shaking for a given oscillator period. | none | 1 |
| 20 | VHR_PGV | VHR is the ratio of peak ground velocity (PGV) to the geometric mean of the two horizontal components, all expressed in natural-log units in the regression. It quantifies how much vertical shaking is expected relative to horizontal shaking for a given oscillator period. | none | 1 |
| 21 | DRVT | DRVT (Duration of a Random Vibration Time-History): A frequency-dependent measure of ground motion duration, representing the time required for a random signal to build up to the observed Fourier amplitude. | s | 1 |